# RESONANT CARBON K-EDGE SOFT X-RAY SCATTERING FROM LATTICE-FREE HELICONICAL MOLECULAR ORDERING: SOFT DILATIVE ELASTICITY OF THE TWIST-BEND LIQUID CRYSTAL PHASE

Chenhui Zhu[1]*, Michael R. Tuchband[2], Anthony Young[1], Min Shuai[2], Alyssa Scarbrough[3], David M. Walba[3], Joseph E. Maclennan[2], Cheng Wang[1], Alexander Hexemer[1]*, Noel A. Clark[2]*

[1]Advanced Light Source, Lawrence Berkeley National Laboratory, Berkeley, CA 94720 USA

[2]Department of Physics and Soft Materials Research Center
University of Colorado Boulder, CO, 80309-0390

[3]Department of Chemistry and Biochemistry and Soft Materials Research Center
University of Colorado Boulder, CO, 80309-0215

ABSTRACT

Resonant x-ray scattering shows that the bulk structure of the twist-bend liquid crystal phase, recently discovered in bent molecular dimers, has spatial periodicity without electron density modulation, indicating a lattice-free heliconical nematic precession of orientation that has helical glide symmetry. *In situ* study of the bulk helix texture for the dimer CB7CB shows an elastically-confined temperature-dependent minimum helix pitch, but a remarkable elastic softness of pitch in response to dilative stresses. Scattering from the helix is not detectable in the higher temperature nematic phase.

The first known thermotropic liquid crystals (LCs), found by Reinitzer in 1888 [1], were the phases of cholesterol derivatives in which nematic ordering produces a well-defined mean local molecular orientation (the director field, *n*(*r*)) that molecular chirality drives to form a helically twisted structure [2]. This chiral nematic helix fills space with uniform twist of *n*(*r*), with *n*(*r*) rotating in a plane, precessing at a constant, spatially homogeneous rate with displacement along a line normal to the plane, the helix (*z*) axis. In 1973, Meyer realized that a homogeneous bend deformation could be added to that of the twist if the helical precession of *n*(*r*) were confined to a cone rather than a plane [3], generating the heliconical nematic (HN) structure, also termed the twist-bend nematic ($N_{TB}$). In 2000, a theoretical proposal by Dozov [4] and simulations by Memmer [5] suggested that even in the absence of molecular chirality the bulk heliconical nematic state might, because of its bent *n*(*r*), be stabilized if made by a fluid of achiral molecules that were suitably bent-shaped. If so, the resulting $N_{TB}$ phase would exhibit fluid, chiral conglomerate domains from achiral molecules.

Several years ago, researchers began to interpret observations on the material CB7CB [4',4'-(heptane-1,7-diyl)bis(([1',1"-biphenyl]-4"-carbonitrile)), [6]] and its homologues, a class of achiral molecular dimers of rigid rods connected by a bent, flexible alkyl spacer, in terms of the heliconical nematic structure [7–13] (Fig 1), with isotropic phase (Iso), nematic phase (N), and the proposed heliconical (HN) nematic structure of the nematic twist-bend ($N_{TB}$) phase appearing vs. temperature *T* in the sequence ISO – (*T* = 113ºC) – N – (*T* = 101ºC) – $N_{TB}$. Phases having $N_{TB}$ characteristics have by now been observed in polar and apolar molecular dimers and trimers [14,15] and bent-core mesogens [16,17]. In spite of these experimental and theoretical [18–21] developments, the structure and nature of the $N_{TB}$ phase is not well understood, with even the proposed heliconical nematic structure still called into question [22,23]. The most direct structural evidence for the heliconical nematic structure to date has been the visualization of periodic nanoscale stripes on freeze-fracture planes in quenched samples with spacings around 80 Å [24,25]. However, NMR [22] and surface structure [23] observations have motivated recent proposals for alternative $N_{TB}$ local structures.

An important direct, *in situ* probe of a bulk structure of an $N_{TB}$ phase would be x-ray diffraction from its periodic planes of distinct orientational ordering. If the twist-bend helix axis is taken to be along *z*, then its director field *n*, giving the local mean molecular orientation, may be written as *n*(*z*) = (sin$\theta$ cos$\varphi$, sin$\theta$ sin$\varphi$, cos$\theta$), where $\varphi$ is the azimuthal angle, given by $\varphi(z) = q_H \cdot z = (2\pi/p_H)z$, where $p_H$ is the nanoscale pitch of the helix and $q_H$ the modulus of the corresponding wavevector. Although periodic, this proposed HN structure has heliconical glide symmetry, under a simultaneous translation $\delta z$ and rotation $\varphi(\delta z)$. It therefore has no electron density modulation (EDM), and is thus not expected to produce diffraction in typical XRD experiments. Thus, the absence of Bragg scattering for hard (10 KeV) x-ray energies [24], is consistent with the notion that the $N_{TB}$ phase is helical.

X-ray diffraction is one of the most useful tools for characterizing the structure of LC phases because of its sensitivity to electron density modulation accompanying positional ordering [26]. However, there are modes of molecular reorientation in LCs that do not produce

EDM, for example the helical precession of the molecular tilt direction around the cone in the chiral smectic C phase of Fig. 1(c)), or the alternation of molecular tilts in anticlinic and clock smectic phases [27]. In these cases, resonant x-ray scattering has been shown to be an effective probe of the ordering [28,29], as the coupling between linearly polarized x-rays and the asymmetric electron cloud of the sample results in a tensorial atomic scattering cross-section for energies near the absorption edge [30], with the scattering contrast dependent on the orientation of the molecule with respect to the polarization direction of the x-ray beam. Such experiments have generally required specially synthesized molecules doped with the target resonant atoms (Cl, S, or P), but recently carbon K-edge scattering has been applied to investigate polymer blends [31], block copolymers [32,33], organic bulk heterojunction solar cells [34] and polymeric transistors [35], in all of which the complex refractive indices of the different components have distinct energy and polarization dependences for x-ray energies near the edge. Motivated by these successes, we recently have shown that otherwise invisible helical ordering in thermotropic LCs can be observed with resonant soft x-ray scattering (RSoXS) at the carbon K-edge [36] by probing the orientation of the carbon bonds in helical nanofilaments formed from bent-core molecules [37].

In this letter, we report the use of RSoXS at the carbon K-edge to provide direct evidence that the $N_{TB}$ phase in CB7CB has a bulk helical periodic modulation of molecular orientation in absence of modulation of electron density. We observe Bragg diffraction, peaked at wavevectors $q_H$, only near the carbon K-edge resonance and only at temperatures $T < 101°C$, the range of the $N_{TB}$ phase and its glassy analogs at lower temperatures [38], indicative of the heliconical structure of the $N_{TB}$ phase, and enabling *in situ* measurement of its bulk helix pitch. The experiments were performed on the soft x-ray scattering beamline (11.0.1.2) at the Advanced Light Source (ALS) of Lawrence Berkeley National Laboratory. The x-ray beam photon energy $E$ was tuned between $E = 250$ eV and $E = 290$ eV in our experiments, a range including the carbon K-edge resonance at $E_R = 283.5$ eV. CB7CB was synthesized as described in the Supplemental Materials (SM) and filled in the isotropic phase between two pieces of 100 nm-thick $Si_3N_4$ membrane (Norcada, Inc.) for transmission powder diffraction study, with the sample cell and beam path in vacuum. The scattering intensity was imaged in two dimensions (2D) by a back-illuminated Princeton PI-MTE CCD, thermoelectrically cooled to -45°C, having a pixel size of 27 μm, positioned 50.6 mm down beam from the sample, and translated off axis to enable recording of diffracted x-ray intensity, $I(\mathbf{q})$ with scattering vector $\mathbf{q} = \mathbf{k}_{scat} - \mathbf{k}_{inc}$ with $q$ the range $q <= 0.08$ Å$^{-1}$ (scattering angle between $\mathbf{k}_{scat}$ and $\mathbf{k}_{inc}$, $\Theta <= 32°$, $2\pi/k_{inc} \sim 44$ Å). The imager then collects arcs of the diffraction rings from the partially unoriented sample. The x-ray beam (300 x 200 μm) is linearly polarized, with a polarization direction that can be rotated continuously from horizontal to vertical.

The RSoXS data from CB7CB are summarized in Figs. 2 and 3 and in the Supplemental Materials, showing that distinct Bragg scattering rings, indicating a periodic bulk lamellar structure at wavevectors in the range $0.06$ Å$^{-1}$ $< q_H <$ $0.08$ Å$^{-1}$ appear in CB7CB for $T < 101°C$ and $E \sim E_R$. At $T = 25°C$ this scattering appears as a well-defined ring at $\Theta_H \sim 32°$, indicative of a lamellar structure of period $p_H = 2\pi/q_H = 80.6$ Å, $q_H = 0.078$ Å$^{-1}$. The 2D arcs $I(\mathbf{q})$, such as shown in Fig. 2(a-f), are azimuthally averaged around the beam center $\mathbf{q} = 0$, over the observable angular range that depends on $q$, to obtain the 1D radial scans $I(q)$ plotted in Fig. 2(g). The detailed structure of the detected rings indicate that the scattering originates from a powder-like mosaic of domains that are internally well-ordered but distributed in azimuthal orientation, size, and peak position. Bragg scattering is detectable for $E$ only in the range 266 eV < $E$ < 287 eV, with a sharp maximum in peak area $A(E)$ at $E_R$ = 283.5 eV, the energy used for the runs vs. temperature ($T$) that gave the temperature data in Fig. 2 and the SM. The peak disappears into the background with increasing $|E - E_R|$ (see SM). For $E > E_R$ this is due in part to the large step in sample attenuation at $E = E_R$ (SM Fig. 2). However, for $E < E_R$ it indicates the decay of the Bragg scattering cross-section, since at lower $E$ the loss of Bragg intensity due to absorption is limited to ~30% or less. Near the N − N$_{TB}$ transition we find $q_H$ ~ 0.064 Å$^{-1}$ ($p_H$ ~ 98 Å), comparable to that measured by FFTEM [24,25] and estimated from electroclinic measurements [39]. These observations indicate that the scattering from CB7CB is due to a helical nematic structure, the only possibility for resonant Bragg diffraction in absence of EDM, and that the scattering cannot be accounted for by proposed local nanostructures that are either not periodic [22] or which have EDM [22,23]. Furthermore, the positive birefringence of the N$_{TB}$ phase, comparable in sign and magnitude to that of the nematic [24,19], indicates that $\theta$ < 54.7°, the magic angle, ruling out any cholesteric-like helix having $\theta$ = 90°, and leaving the heliconical nematic as the only remaining possible structure. In this case, since the geometries for $\varphi = \pi$ and $\varphi = 2\pi$ are distinct, the lowest order resonant Bragg scattering is at $q_H = 2\pi/p_H$, where $p_H$ is the full helix pitch, the distance along **z** for a $\varphi = 2\pi$ rotation around the cone (Fig. 1(d)).

The RSoXS technique has enabled the first *in situ* studies of the bulk heliconical nematic structure and pitch versus temperature, with results of a heating-cooling-reheating cycle (between $T$ = 25ºC and the nematic phase) shown in Figs. 2 and 3 (see SM). This series of measurements started from $T$ = 25ºC with a sample previously melted into the Iso phase and cooled to room temperature in order to fill the cell. For the heating scans $I(\boldsymbol{q},T)$ at low $T$ is generally a single ring as in Fig. 2(a), but, remarkably, broadens with increasing $T$ into a pattern of distinct arcs, each a partial ring localized in $q$ with a range of azimuthal orientations, as shown in Fig. 2(c,d).

At higher $T$ in the N$_{TB}$ range the scattering is a superposition of distinct peaks at different $q_H$ values (Fig. 2(c-e)), some quite sharp. The half-width at half maximum (HWHM) of the narrowest of these is found to be $\delta q_H$ ~ 2*10$^{-4}$ Å$^{-1}$. This width is comparable to the intrinsic resolution of the scattering geometry, $\delta q_{res} = k_{inc}\cos(\Theta/2)\cos^2\Theta(w/L)$ ~ 2*10$^{-4}$ Å$^{-1}$ HWHM, limited by the width of the beam, $w$ ~ 200 μm, and $L$ = 50.6 mm is the sample to detector separation, indicating that the scattering from single domains can be nearly resolution limited, even for $T$ close to the N−N$_{TB}$ transition. Since $\delta q_{res}$ corresponds to coherence length in the scattering of $l = \delta q_{res}^{-1}$ ~ 1 μm, the single peak width shows that the domains can be locally well-ordered over micron scale volumes, corresponding to coherent ordering over > 100 periods of the periodic lamellar structure, consistent with the FFTEM images of micron-scale areas [24,25]. The broader peaks may indicate local order with a more limited range of layer correlation or be a superposition of domains of differing peak position.

The spread in peak positions at a given $T$ implies a distribution of pitches $P(p_H)$, characterized in Fig. 2(g,h) by plotting the higher and lower limits of pitch $p_H$ at each temperature for the heating and cooling scans. The width of the distribution of pitches, $\Delta p_H$, is narrowest at low $T$, where $\Delta p_H/p_H \sim 0.35/80 = 0.0044$, and increases to a maximum of $\Delta p_H/p_H \sim 8.6/93 = 0.093$ as the transition to the nematic is approached. The width $\Delta p_H$ then decreases over the phase coexistence range of a few degrees near the transition (vertical cyan bar in Figs. 3(d-f)), as the smaller peaks transition into the nematic, completed by the disappearance of the last single peak to give no detectable Bragg scattering in the nematic phase, a behavior indicative of a first-order transition.

Plots of the individual temperature scans in Fig. 3(a-c) show the distinct broadening of the pitch distribution with increasing $T$ and upon cooling the corresponding shrinking to a single peak or a narrow distribution of peaks. In each case, the variation in the value of the higher $p_H$ limit, $p_H(T)_{high}$, is much larger and more erratic than that of the lower limit, $p_H(T)_{low}$. Superposition of the higher and lower limit data in Figs. 3(d-f) shows that, in fact, $p_H(T)_{low}$ exhibits very little variation among the three runs, the principal deviation among them being a shift of $p_H(T)$ to lower temperatures by ~1ºC relative to the heating curves near the N−N$_{TB}$ transition, due the hysteresis of the transition on cooling vs. heating. This constancy of $p_H(T)_{low}$, also evident in typical families of multi-peak structures obtained by taking series of azimuthal averages over narrow azimuthal sectors (see SM), can be taken as evidence that $p_H(T)_{low}$ is, in fact, the strain-free pitch of the N$_{TB}$ in CB7CB. In CB7CB on heating, the strong layer expansion must mean that the layering system is under varying degrees of local compressive stress. While layering domains with varying degrees of expansion appear, there are none showing up with the layering substantially compressed. This implies that the elastic energy required for a change of pitch is very asymmetric at higher temperatures in the N$_{TB}$ phase, with the stress required for a certain fractional dilation being much smaller than that for compression. An additional feature to be pointed out is that during the cooling run the N$_{TB}$ appears at T ~ 100 ºC as a single peak located on the lower limit curve, a further indication that the lower limit curve is giving nearly strain-free pitch values since isolated N$_{TB}$ domains are likely to be strain free.

The broad distribution $P(p_H)$ of sharp peaks indicates the presence of domains that have homogeneous internal strain induced by an inhomogeneous distribution of varying, local stresses. An estimate of the form of the elastic energy for change of pitch, $U(p_H)$, has been made by assuming that these local stresses are randomly distributed and that $U(p_H) \propto -C\ln[P(p_H)]$, where $C$ is an unknown measure of the RMS stress fluctuation (see SM). The softening of $U(p_H)$ for layer dilation at high $T$ is evident. Many of these peaks in $I(q)$ come from rings in $I(q)$ that extend in the images over many tens of degrees of azimuthal angle $\phi$, indicative of structures having curved layers with a certain pattern of pitch dilation everywhere, perhaps in focal conic domains, which are commonly seen in the twist-bend nematic [12]. However, optical microscopy study of such structures shows that, once formed, they tend to persist upon cooling and are thus likely present at the lower temperatures. In this case the collapse of the distribution of $p_H$ to a narrow range of values would have to be due to elastic resistance to layer dilation that increases with decreasing temperature.

ASSOCIATED CONTENT
Supplemental Materials.

AUTHOR INFORMATION
Corresponding Author
Email: chenhuizhu@lbl.gov, ahexemer@lbl.gov, noel.clark@colorado.edu

NOTES
The authors declare no competing financial interest.

ACKNOWLEDGMENT
We acknowledge use of beamline 11.0.1.2 of the Advanced Light Source supported by the Director of the Office of Science, Office of Basic Energy Sciences, of the U.S. Department of Energy under contract no. DE-AC02-05CH11231. This work was supported by the Soft Materials Research Center NSF MRSEC award DMR-1420736.

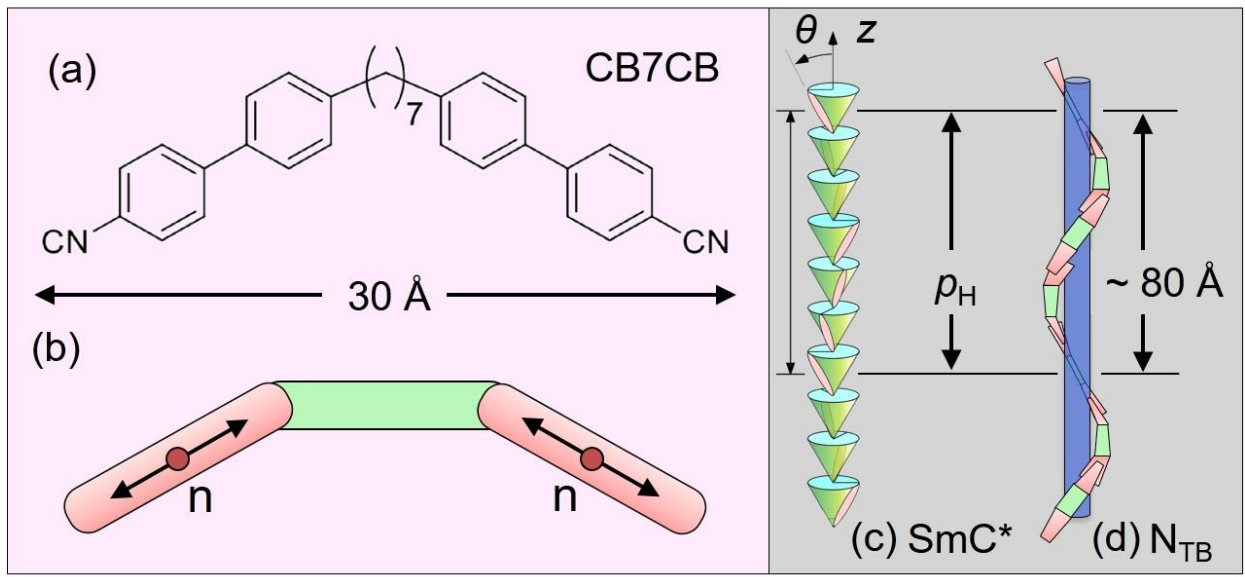

FIG. 1. (a) The dimer CB7CB consists of two cyano-biphenyl molecular arms tethered by a 7 carbon alkyl chain. (b) The molecule can be modeled as two rigid-rod segments (red) tied together by a flexible linker (green). (c) Conical helix of a chiral smectic C layered phase. (d) Proposed heliconical structure of the twist-bend ($N_{TB}$) phase. This molecular arrangement has helical glide symmetry and therefore no electron density modulation. It Bragg scatters x-rays only near an absorption edge resonance of a constituent atom, where the cross-section becomes dependent on molecular orientation. In both (c) and (d), the rigid molecular components make an average angle $\theta$ with respect to the helix axis, $z$. $P$ is the polarization accompanying the heliconical structure.

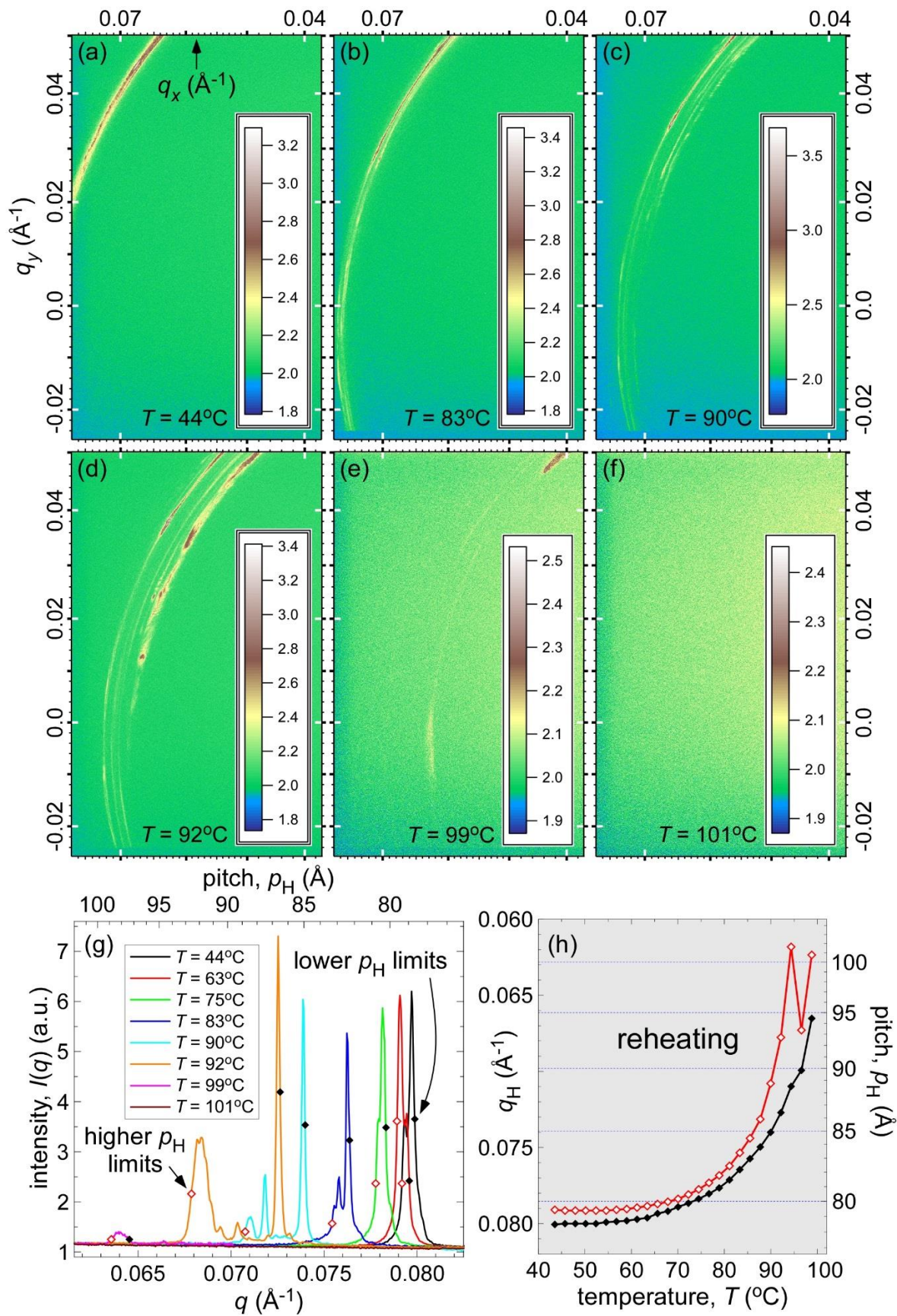

FIG. 2. (a-f) RSoXS, with incident x-ray photon energy at the carbon K edge ($E = 283.5$ eV), of the twist-bend nematic phase of CB7CB as a function of temperature on reheating, after heating into the nematic and cooling to $T = 25$ °C. The scattering arcs generally broaden and shift to smaller $q$ as $T$ increases, corresponding to a mosaic of $N_{TB}$ domains with a variety of pitch lengths that reach up to ~100Å at temperatures near the N−$N_{TB}$ transition. (f) There is no observable scattering in the higher temperature nematic phase. (g) Radial scans in $q$ of azimuthal averages of $I(\mathbf{q})$ about the beam center $\mathbf{q} = 0$ of the images in (a-f), each for the entire available range of $\varphi$. The limits of the pitch distribution are measured from the half maximum of the outermost scattering peaks, with the upper $p_H$ limit denoted by the open red diamonds and the lower $p_H$ limit with closed black diamonds. (h) The higher and lower limits of $q_H$ and $p_H$ as measured from line scans which include the entire scattering arc. At high temperatures the trend in the lower limit of $p_H$ is significantly smoother than the higher limit, implying that the lower $p_H$ limit represents the strain-free pitch of CB7CB.

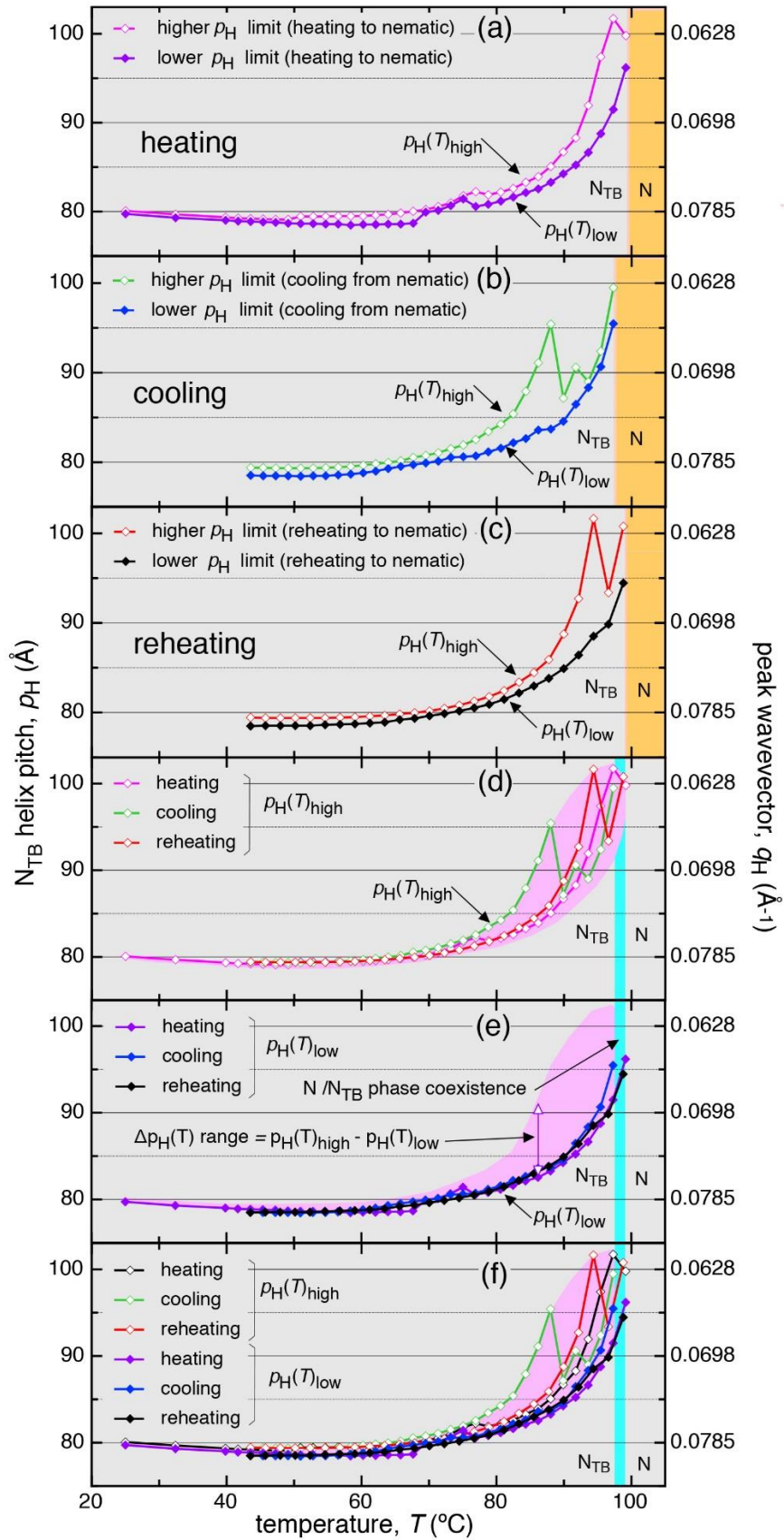

FIG. 3. (a-c) CB7CB pitch range limits as a function of temperature on initial heating (a), cooling (b), and reheating (c), measured from Fig. 2 and SM Figs. 3 and 4 as described in the caption of Fig. 2. The orange shaded region indicates the temperature interval of the nematic phase, where no RSoXS is observed, and shows hysteresis upon heating vs. cooling. (d-f) Plots combining the higher (d), lower (e), and all (f) $p_H$ limit data from the three temperature scans. The purple-pink shading shows $\Delta p_H(T) = p_H(T)_{high} - p_H(T)_{low}$, the range of measured pitch lengths indicated by the data. The width of the vertical cyan bar denotes the temperature interval over which there is $N_{TB}/N$ phase coexistence.

Chenhui Zhu[1]\*, Michael R. Tuchband[2], Anthony Young[1], Min Shuai[2], Alyssa Scarbrough[3], David M. Walba[3], Joseph Maclennan[2], Cheng Wang[1], Alexander Hexemer[1]\*, Noel Clark[2]\*

[1]*Advanced Light Source, Lawrence Berkeley National Laboratory, Berkeley, CA 94720 USA*

[2]*Department of Physics and Soft Materials Research Center
University of Colorado Boulder, CO, 80309*

[3]*Department of Chemistry and Biochemistry and Soft Materials Research Center
University of Colorado Boulder, CO, 80309*

**SM FIG. 1**: Bragg scattering from the N$_{TB}$ phase of CB7CB at T = 25 ºC for x-ray energies near the carbon K-edge.

**SM FIG. 2**: Attenuation length of 5CB as measured in the range 250 eV to 290 eV.

**SM FIG. 3**: RSoXS, with incident x-ray photon energy at the carbon K edge (E = 283.5 eV), of the twist-bend nematic phase of CB7CB as a function of temperature on a heating cycle.

**SM FIG. 4**: RSoXS, with incident x-ray photon energy at the carbon K edge (E = 283.5 eV), of the twist-bend nematic phase of CB7CB as a function of temperature on cooling from the nematic.

**SM FIG. 5**: RSoXS, with incident x-ray photon energy at the carbon K edge (E = 283.5 eV), of CB7CB just below the N−N$_{TB}$ transition in neighboring 10° sectors.

**SM FIG. 6**: Mutual comparison of three distinct measurements of the helix pitch of the N$_{TB}$ phase of CB7CB.

**SM FIG. 7**: Estimate of the pitch elastic energy function $U(p_H)$ for CB7CB in several temperature regimes of the N$_{TB}$ phase of CB7CB.

**Synthesis of CB7CB**

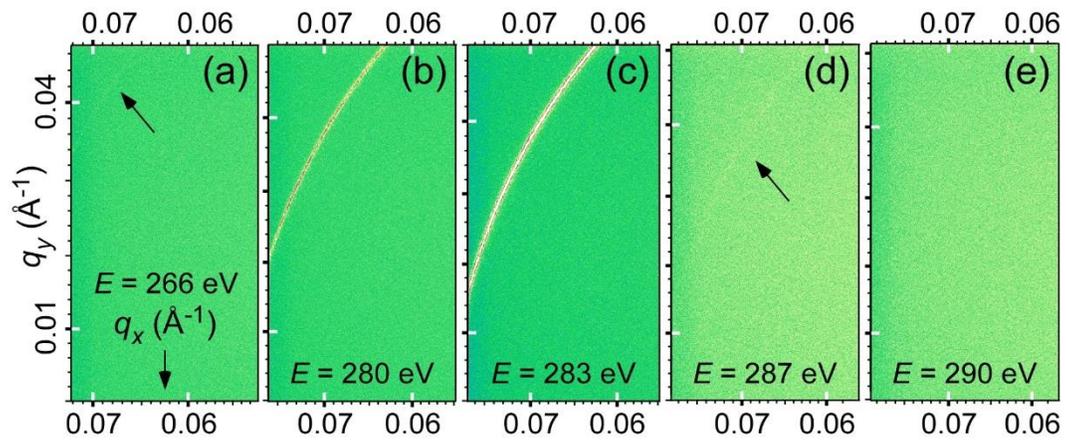

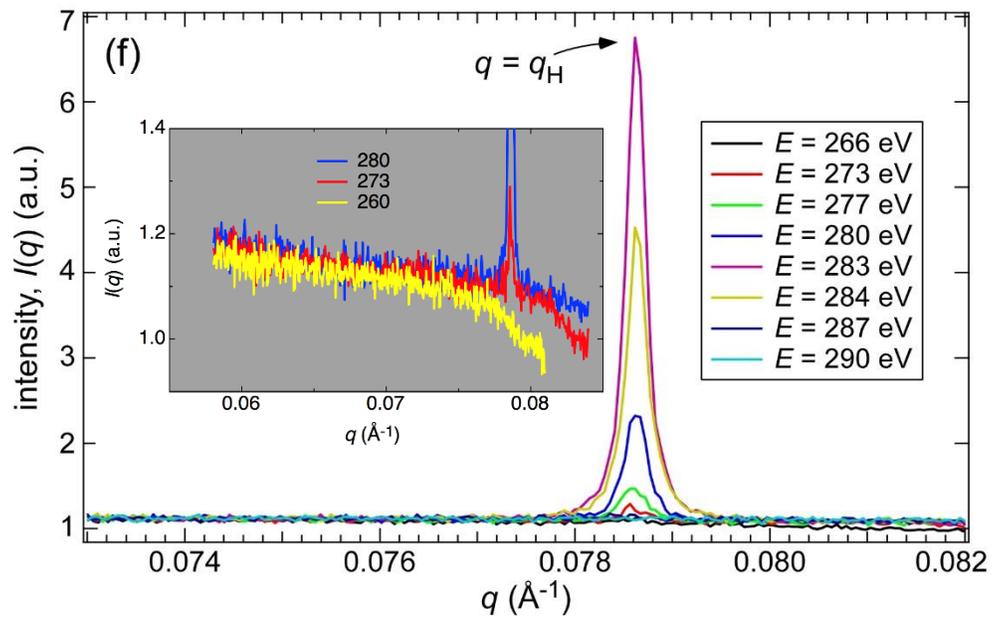

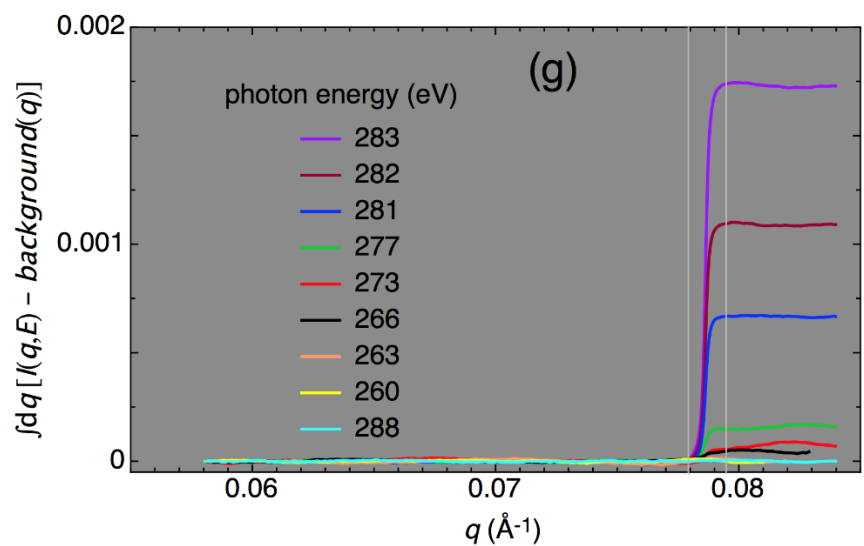

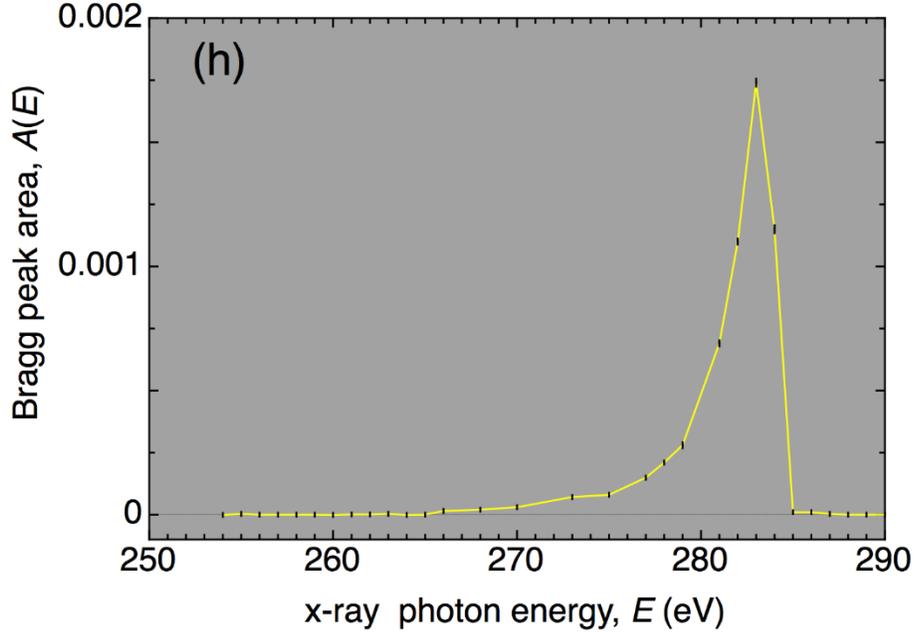

**SM FIG. 1**: Bragg scattering from the $N_{TB}$ phase of CB7CB at T = 25 ºC for x-ray energies near the carbon K-edge. (a-e) 2D x-ray diffraction images with the incident x-ray photon energy, $E$, tuned to several values. The peak intensity increases as the incident x-ray photon energy approaches the carbon K-edge resonance at $E_R \sim 283.5$ eV, and subsequently decreases for $E > E_R$. (f) Radial scans in $q$ of azimuthal averages of $I(\mathbf{q})$ about $\mathbf{q} = 0$ at several beam energies. The inset shows selected scans on an expanded vertical scale. (g) Integration over $q$ of background-subtracted $I(q,E)$ over the range $(0.058,q)$ to obtain area of the peaks, $A(E)$, equal to the step between the vertical lines. The vertical dimension of the plotted symbols represents the uncertainty in their measurement. (h) Bragg peak area $A(E)$ vs x-ray beam energy for $E \sim E_R$. The $N_{TB}$ phase of CB7CB exhibits significant contrast for scattering at energies $E \sim E_R$, but which is not observable for $E < 266$ eV, evidence for the helical glide symmetry of the heliconical nematic structure.

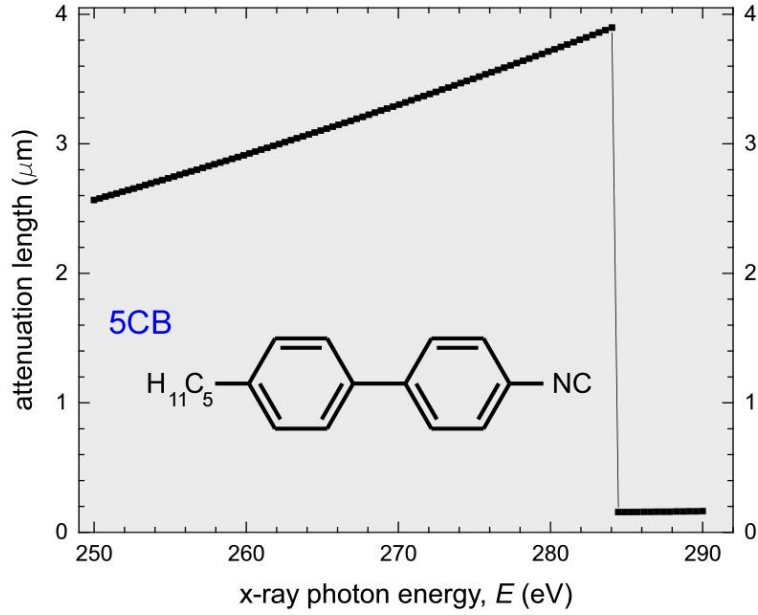

**SM FIG. 2**: Attenuation length of 5CB as a function of incident x-ray energy, $E$, in the range 250 eV $< E <$ 290 eV, showing the carbon K absorption edge at $E_R = 283.5$ eV. Because of the structural similarity of 5CB and CB7CB, the attenuation length of the two systems should be essentially identical. The decrease in the measured scattered intensity in CB7CB when tuning with $E > E_R$ (see SM Fig. 1(h)) is made significantly sharper by the abrupt decrease in the attenuation length for $E > E_R$. For $E < E_R$, the attenuation length is comparable to the sample thickness.

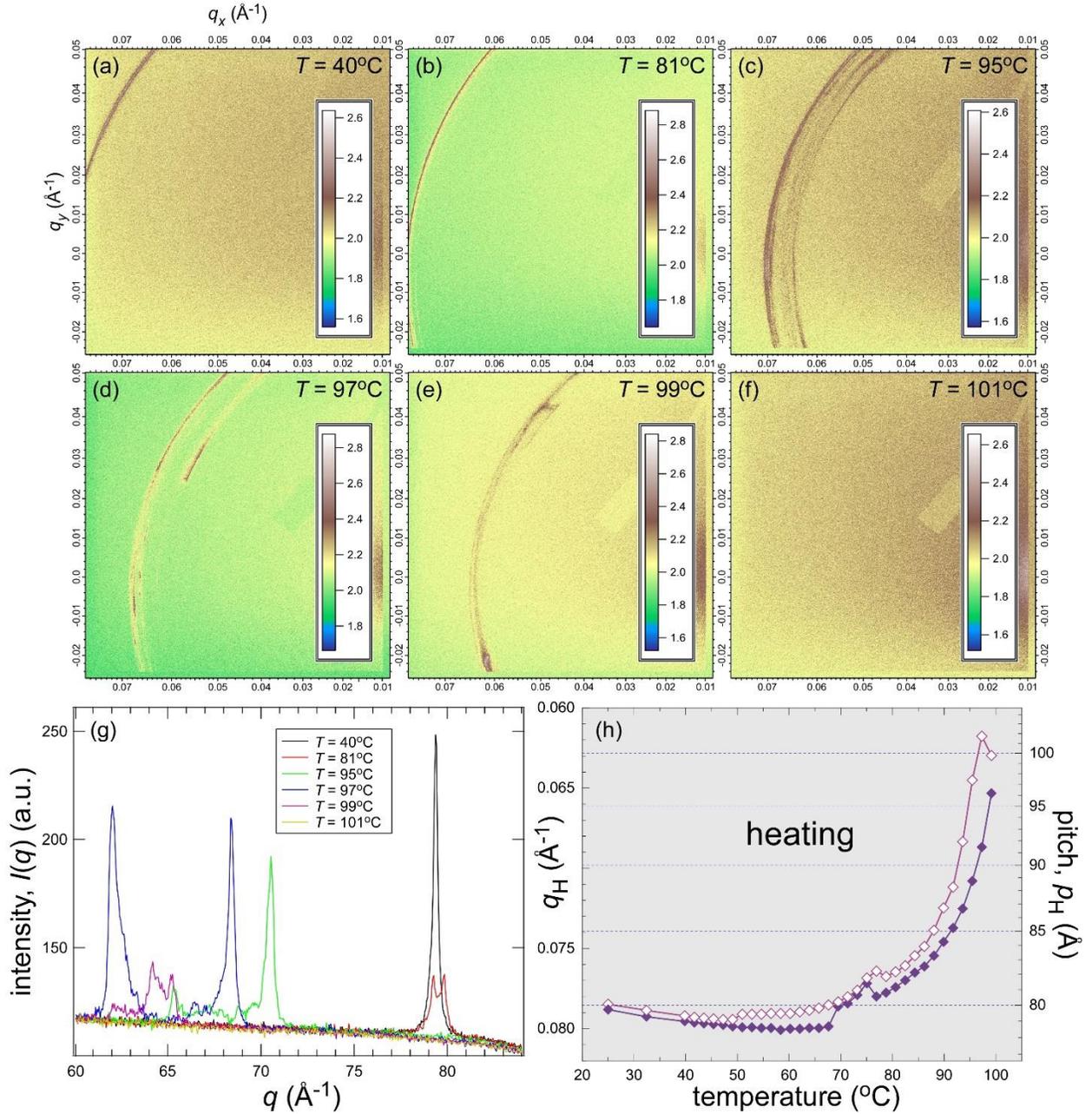

**SM FIG. 3**: RSoXS, with incident x-ray photon energy at the carbon K edge (E = 283.5 eV), of the twist-bend nematic phase of CB7CB as a function of temperature in a heating cycle. (a-e) 2D x-ray scattering images on increasing temperature. The initial scattering arc at low temperature (a,b) splits into a multitude of sharp scattering arcs at higher temperatures (c-e), with the arcs shifting toward smaller $q$-values. No scattering is visible in the nematic phase (f). (g) Radial scans in $q$ of azimuthal averages of $I(\mathbf{q})$ about $\mathbf{q} = 0$ of the images in (a-f) over the entire available range of $\phi$. (h) The higher and lower limits of $p_H$ during this temperature scan, obtained as described in Fig. 2.

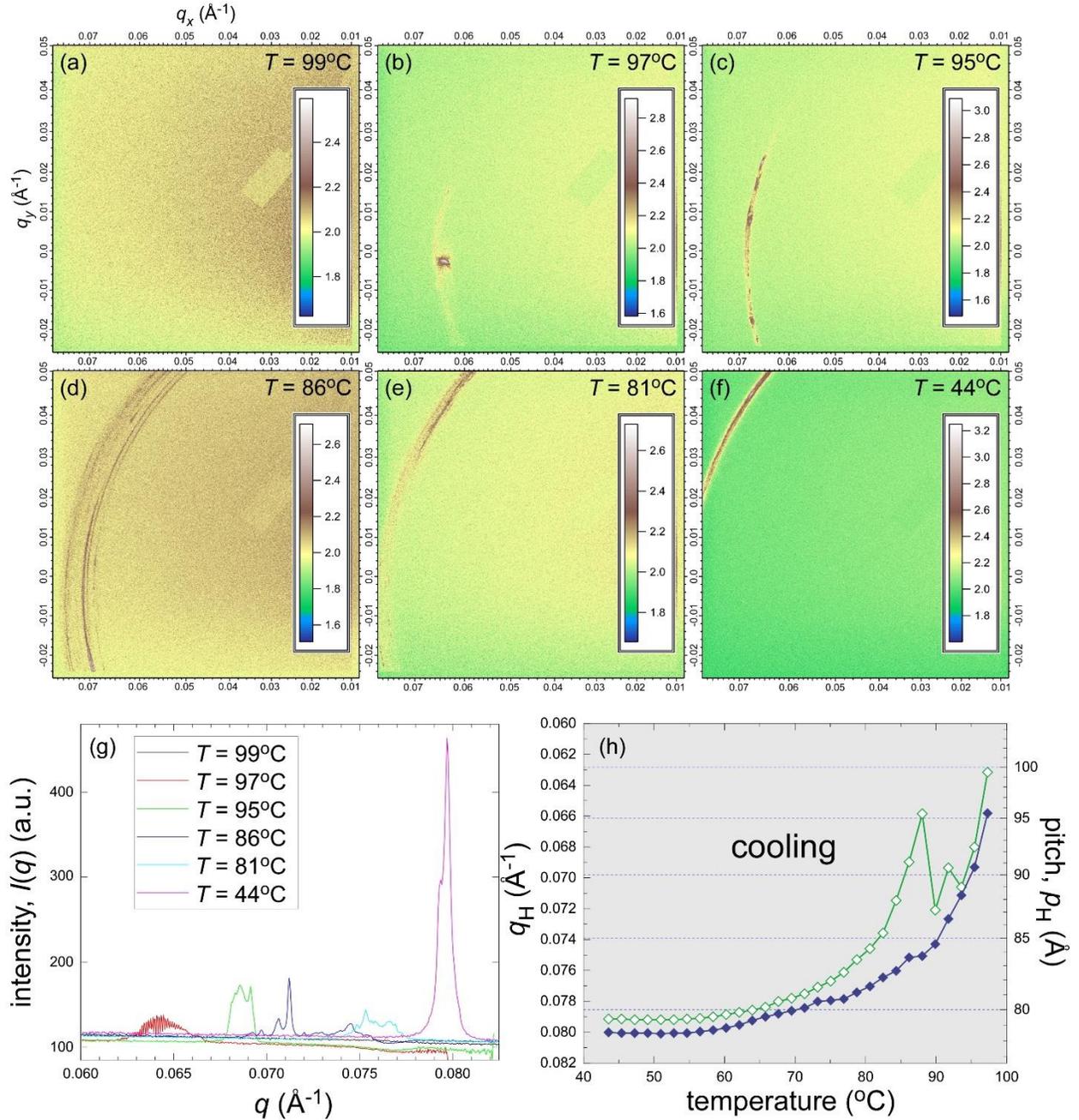

**SM FIG. 4**: RSoXS, with incident x-ray photon energy at the carbon K edge (E = 283.5 eV), of the twist-bend nematic phase of CB7CB as a function of temperature on cooling from the nematic. (a-f) 2D x-ray scattering images on cooling. (a) No scattering is visible in the nematic phase. Small scattering spots and arcs appear on cooling from the nematic (b,c), indicating the nucleation of $N_{TB}$ domains. On further cooling, we observe a broad distribution of sharp scattering arcs (d), denoting the presence of many large-scale $N_{TB}$ domains with a variety of pitch lengths. On further cooling, the arcs converge and shift toward higher $q$ (e,f). This behavior is nearly identical to that observed on heating (Fig. 2 and SM Fig. 3). (g) Radial scans in q of azimuthal averages of $I(\mathbf{q})$ about $\mathbf{q} = 0$ of the images in (a-f) over the entire available range of $\phi$. (h) The higher and lower limits of $p_H$ during this temperature scan, obtained as described in Fig. 2.

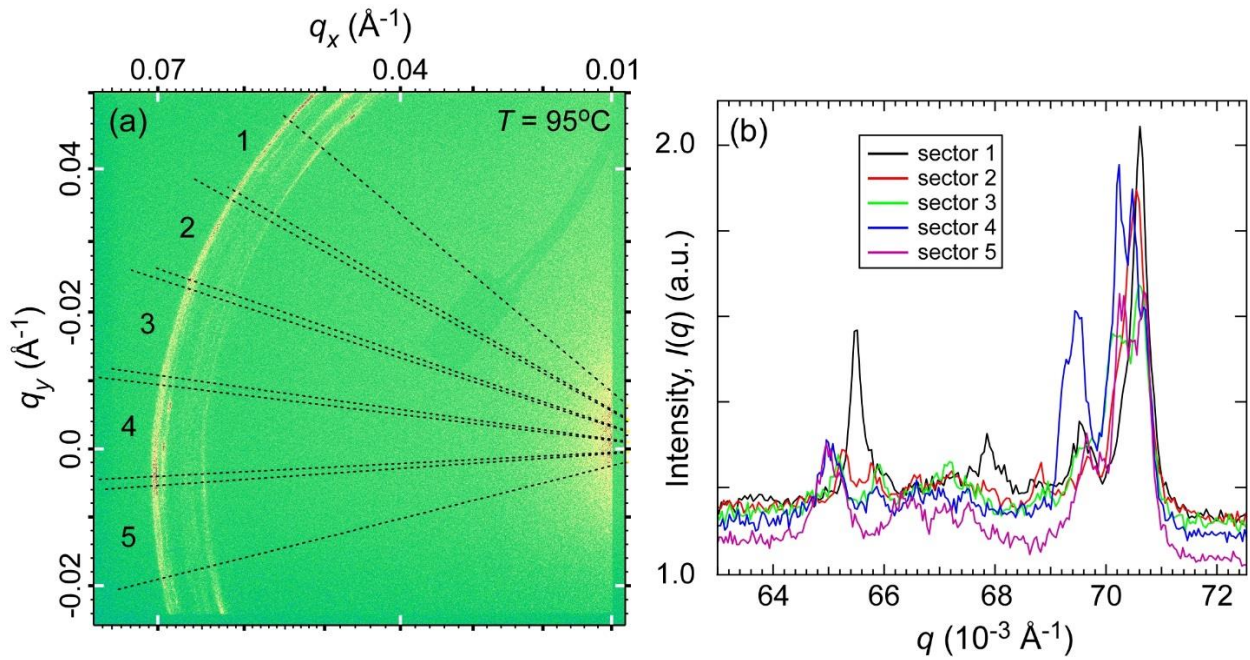

**SM FIG. 5**: RSoXS, with incident x-ray photon energy at the carbon K edge (E = 283.5 eV), of CB7CB just below the N−N$_{TB}$ transition in neighboring 10° sectors. (a) 2D x-ray scattering image of CB7CB showing five sectors which are separately integrated in 10° azimuthal intervals to give the plot in (b). The black dashed lines indicate the boundaries of the sectors.

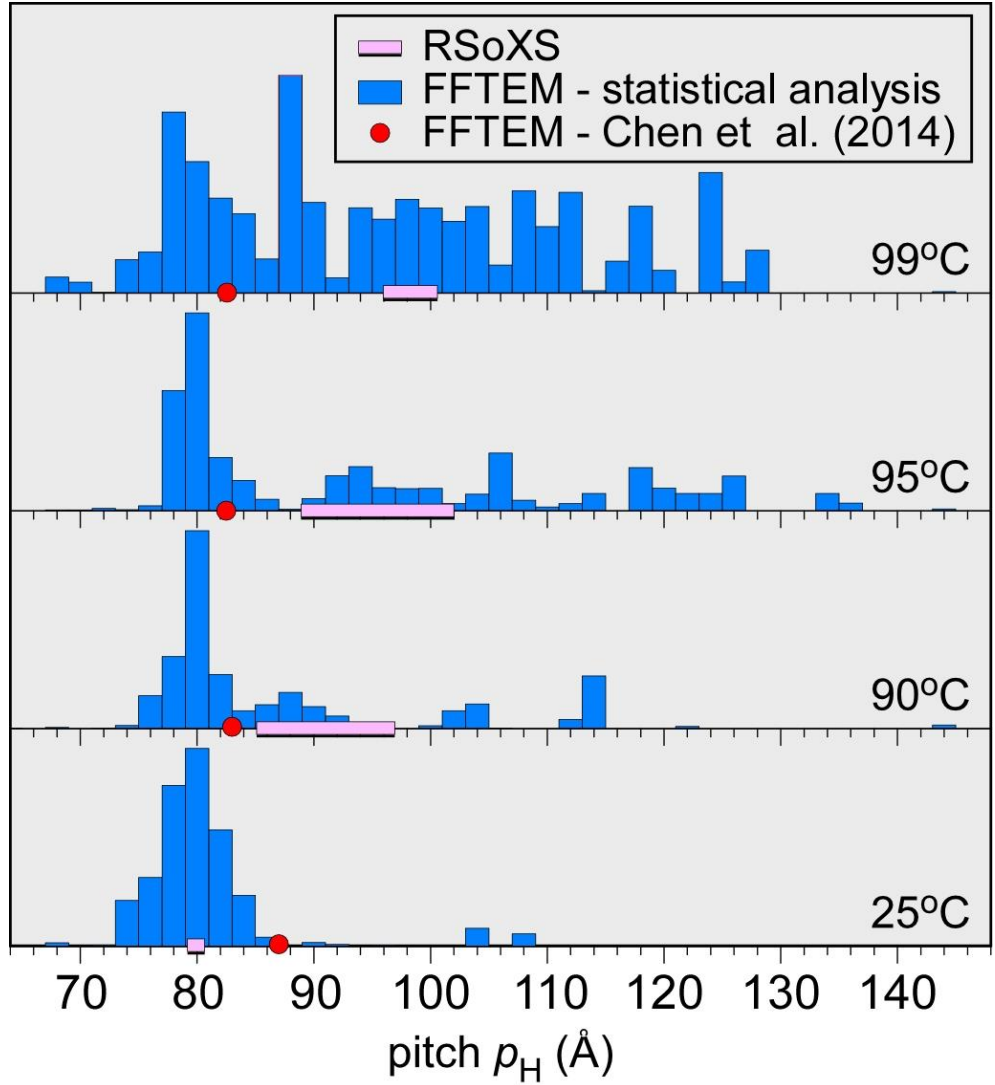

**SM FIG. 6**: Mutual comparison of three distinct measurements of the helix pitch of the $N_{TB}$ phase of CB7CB. The horizontal purple-pink/black bars show the pitch ranges obtained in the RSoXS data of Fig. 3, $\Delta p_H = p_H(T)_{high} - p_H(T)_{low}$, as indicated by the purple-pink shaded region in Figs. 3(d-f). These are compared with two sets of freeze-fracture transmission electron microscopy (FFTEM) data. In the FFTEM measurements, the CB7CB sample is quickly quenched from the indicated temperature to liquid propane temperature, fractured, and a platinum-carbon replica of the fracture surface topography prepared for imaging in the TEM. The red circles are FFTEM results of Chen et al. [24], based on several FFTEM images at each temperature. The blue histograms were obtained from a much more extensive set of FFTEM images of $N_{TB}$ monodomains [41], measuring the pitch length in each $N_{TB}$ domain in a given image and weighting it by the area over which it was observed. The histograms indicate the 'frequency that a given pitch length was observed by FFTEM at the given quenching temperature $T$. In addition to the broader distribution of layer spacing at high temperature, the apparent distribution of pitch is broadened in the FFTEM method by both having fracture planes that are not normal to the helix axis and having the image of fracture planes not normal to the electron beam direction [41]. At high temperature Chen et al. observed principally focal conic domains, for which these geometrical effects are reduced. Both the RSoXS and FFTEM measurements indicate increased softness for dilation of the pitch with increasing temperature.

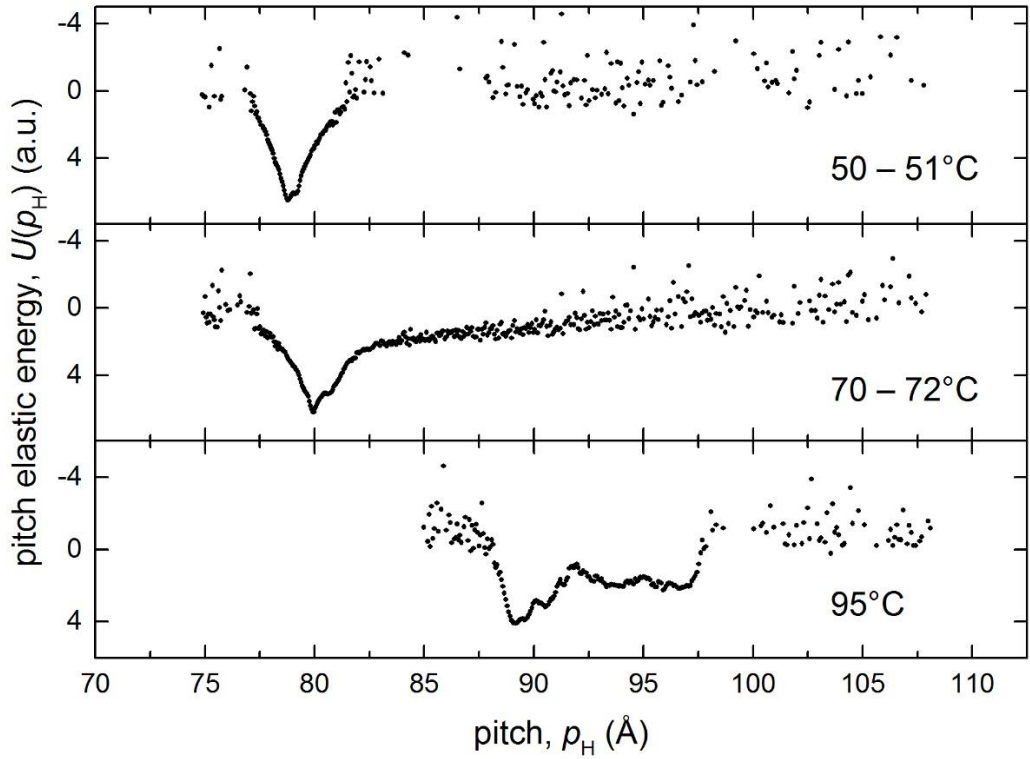

**SM FIG. 7**: Estimate of the pitch elastic energy function $U(p_H)$ for CB7CB in several temperature regimes of the $N_{TB}$ phase of CB7CB. $U(p_H)$ is obtained under the assumption of a distribution $P(p_H)$ of domains of internally uniform $p_H$, generated by a distribution of random stresses tending to change the pitch, and calculated as $U(p_H) \propto -C\ln(P(p_H))$, where $C$ is an unknown constant proportional to the mean square local stress.

## Synthesis of CB7CB:

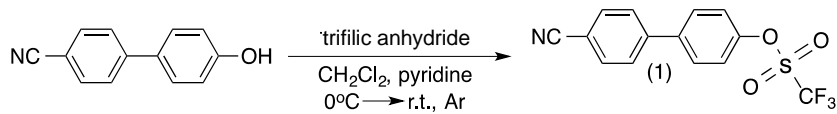

**1,1,1-trifluoro Methanesulfonic acid-4'-cyano[1,1'-biphenyl]-4-yl ester (1)**

5 g (25.6 mmol) 4-cyano-4'-hydroxybiphenyl was dissolved in 85 mL of $CH_2Cl_2$, cooled to 0°C and placed under argon. 32 mL (32.0 mmol) of 1 M triflic anhydride in $CH_2Cl_2$ was added drop wise. 3.5 mL (43.5 mmol) of anhydrous pyridine was then added via syringe and reaction warmed to room temperature over night. After 20 hours, the reaction mixture was poured over ice, diluted with $CH_2Cl_2$, washed once with water, once with 3% $H_2SO_4$, once with saturated aqueous NaCl, dried over both $MgSO_4$ and placed under reduced pressure to afford 8.11 g (24.8 mmol, 97% crude yield) light orange solid.

$^1$H NMR (400 MHz, $CDCl_3$): 7.80-7.72 (m, 2H), 7.71-7.62 (m, 4H), 7.45-7.36 (m, 2H).
$^{13}$C NMR (101 MHz, "$CDCl_3$"): 149.69, 143.61, 139.58, 132.78, 129.10, 127.83, 122.08, 118.71 (q, $CF_3$), 118.53, 111.85, 77.31, 76.99, 76.83, 76.67.

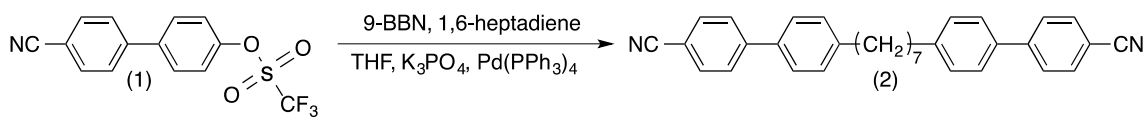

**4',4'-(heptane-1,7-diyl)bis(([1',1''-biphenyl]-4''-carbonitrile)) (2)**

1.18 g (12.2 mmol) 1,6-heptadiene was dissolved in 20 mL dry THF and placed under argon. 51 mL (26.0 mmol) 0.5 M 9-Borabicyclo(3.3.1)nonane (9-BBN) in THF was added to the flask and the mixture refluxed for 3 hours. To a second flask, 7.85 g $K_3PO_4$ (37.0 mmol), 6.89 g **(1)** (21.1 mmol), and 40 mL THF were added and sparged with argon for 1 hour. 815 mg $Pd(PPh_3)_4$ (.710 mmol) dissolved in 20 mL THF was cannulated into the second flask, followed by the 9-BBN adduct from flask 1. The contents of flask 2 were refluxed for 3 days under argon. THF was removed under reduced pressure and the crude oil was subsequently dissolved in $CH_2Cl_2$. The solution was washed twice with water, and once with saturated aqueous NaCl, dried over $MgSO_4$, and placed under reduced pressure to afford a dark brown oil. The crude oil was purified via flash chromatography with an eluent gradient of 95:5:2 hexanes:ethyl acetate:$CH_2Cl_2$ to 80:20:2 hexanes:ethyl acetate:$CH_2Cl_2$, then recrystallized multiple times in hexanes and acetonitrile to afford 2.11 g (4.64 mmol, 38% yield) white solid.

$^1$H NMR (300 MHz, $CD_2Cl_2$): 7.77-7.64 (m, 7H), 7.59-7.48 (m, 4H), 7.35-7.24 (m, 4H), 2.71-2.60 (m, 4H), 1.65 (q, $J$ = 7.4 Hz, 4H), 1.42-1.30 (m, 6H).
$^{13}$C NMR (75 MHz, $CDCl_3$): 145.69, 143.80, 136.61, 132.68, 129.29, 127.58, 127.20, 119.15, 110.68, 35.73, 31.48, 29.47, 29.33.